\documentclass[pra,twocolumn,showpacs,showkeys]{revtex4}

\usepackage{amssymb, amsmath, amsfonts, latexsym, verbatim, mathrsfs}

\usepackage{graphicx}

\newtheorem{theorem}{Theorem}
\newtheorem{itdefinition}{Definition}
\newenvironment{definition}{\begin{itdefinition}}{\end{itdefinition}}

\begin{document}

\title{Quantum control of noisy channels}

\author{Raffaele Romano}


\author{Peter van Loock}

\affiliation{Max Planck Research Group, Institute of Optics,
Information and Photonics, University of Erlangen-N\"{u}rnberg}


\begin{abstract}


\noindent Sender and receiver can control noisy channels by means of the resources they own, that is local operations, potentially correlated using classical communication, and entangled pairs shared between them. Using the Choi-Jamiolkowski isomorphism, we express the control of a noisy channel in terms of control of (non-normalized) states, and show how the available resources enter the analysis. Our formalism provides a general scheme for the protection of quantum state transmission when a single use of the physical channel is considered. Moreover, it paves the way to the application of control theoretical tools to the study of noisy channels. We define the notion of complexity of a noisy channel, as a measure of how demanding is to engineer specific manipulations of a channel. We provide some examples of both deterministic and probabilistic protocols leading to a decreased complexity.

\end{abstract}



\maketitle


\section{Introduction}\label{sec1}

The properties of the microscopic world suggests that isolated quantum systems
are good candidates for the implementation of outreaching technologies in the
context of information transmission and processing~\cite{niel}. The ideal manipulation
of a quantum system is described by a unitary transformation $\varepsilon_0$,
acting on the state $\rho$ of the system as
\begin{equation}\label{chan0}
    \rho^B_0 = \varepsilon_0 [\rho^A],
\end{equation}
where $A$ (and $B$) denote the state before (respectively after) the transformation.
$\varepsilon_0$ is the identity ${\mathcal I}$ when we want to store or
transmit the state without modifying it, or a unitary transformation,
when we want to process the information contained in the state by means of
a quantum gate.

The main difficulty for the implementation of quantum technologies is that quantum
systems are fragile, since they interact with the surrounding environment.
Therefore, in full generality, real quantum systems are subject to an irreversible
evolution, described by a completely positive map $\varepsilon$,
\begin{equation}\label{chan}
    \rho^B = \varepsilon [\rho^A],
\end{equation}
where $\varepsilon$ deviates from the ideal operation $\varepsilon_0$, and
reduces to it in the absence of environmental interaction.
Although many concepts and ideas developed in this work can be
applied to the general scenario, for simplicity we refer to the transmission of a quantum state from
a sender to a receiver. Therefore, we call $\varepsilon$ a {\it quantum channel},
and the corresponding ideal operation will be the identity. $A$ represents
the sender of the quantum state, and $B$ the receiver.

There are several quantities that can express the impact of the noise on the transmission,
and the departure of $\varepsilon$ from $\varepsilon_0 = {\mathcal I}$. However,
the sender and the receiver can affect the channel by performing some manipulations
on the states $\rho^A$, and $\rho^B$ respectively. Standard actions are: enlargement of the system by
means of auxiliary systems (ancillae), reduction of it by discarding some degrees of freedom,
application of local unitary operations or measurements on system and ancillae, and, finally,
transmission of classical information about the measurement outcomes for further
processing ({\it feedforward}). These additional resources provide some control
over the noise, since they can be used to counteract the irreversibility induced by
the environment. Accordingly, $\varepsilon$ is replaced by a new channel
$\tilde{\varepsilon} = \tilde{\varepsilon} (\varepsilon, u)$, where $u$ represents the control
parameters, entering through the manipulations performed by $A$ and $B$~\footnote{
In this paper we shall always assume that these manipulations are performed
instantaneously. This leads to constant controls $u$. However, in a more general scenario
it should be appropriate to deal with time-dependent controls $u(t)$, as pointed out in
the Conclusions of the paper.}. The form of $\tilde{\varepsilon}$ is strongly dependent
on the resources that sender and receiver can use.

In the past decade, several strategies for efficiently suppressing the decohering
action of the environment have been proposed, relying on different resources.
For the reliable transmission of quantum states, these techniques
fall into at least two categories: Quantum Error Correcting Codes (QECC)~\cite{shor,stea,cald,benn,knil,niel}
and protocols based on Quantum Teleportation (QT)~\cite{benn2}.

In QECC, the quantum state to be transmitted is encoded in a larger composite system, whose elementary
constituents are sent through independent channels. Finally, the errors on not too
many of the individual constituents can be detected and the logical information restored. The needed resources are
local ancillae to be coupled to the original system, local unitary operations, encoding the logical
information into the code space, and providing the error detection and correction, and finally
the ability to independently use the physical channel many times (or rather to use many
independent channels in parallel). In QT, the sender performs a Bell measurement involving the
system to be transmitted and his part of a (near-) maximally entangled pair, shared with the receiver. Then,
he sends the measurement outcome to the receiver using noiseless classical communication, and
the receiver applies a suitable unitary transformation on his own part of the entangled pair,
reproducing the unknown quantum state to be transmitted.

Apparently, QT schemes are completely different from QECC, since they ask for the
existence of a maximally entangled pair shared between sender and receiver,
together with one-way classical communication between the two parties. However, it has been proven
that QECC and QT are equivalent for a significant class of quantum channels, the so-called {\it
Pauli-diagonal channels}~\cite{benn}. The key-idea is that sender and receiver must use the noisy channel,
to transmit half of the pairs, and then distill the corrupted states
to obtain the needed maximally entangled pairs, by using purification protocols~\cite{benn3,dur2}.
In this perspective, it is assumed that the necessary entanglement is not a prerequisite, but it
is shared through the transmission process. Therefore, quantum communication needs the same
local resources whether it is based on QECC or QT.


However, entanglement can also be seen as an initial resource, prepared off-line
before the use of the channel. For instance, it is possible to improve the quality of entanglement
by means of distillation protocols, in which a reduced number of highly entangled pairs
can be obtained by a larger number of poorly entangled ones by means of local operations and classical
communication. Initially shared entanglement is a useful resource for improving the transmission of
both classical and quantum information. Under this perspective, in {\it superdense coding} the classical
capacity of a quantum channel is doubled if sender and receiver share a maximally entangled
pair~\cite{benn4}. Analogously, it has been proven that QECC can be generalized
if an even partial entanglement is available to sender and receiver at the beginning~\cite{brun}.
The corresponding codes, called Entanglement Assisted QECC (EAQECC),
perform better than standard QECC, since they require a reduced number of ancillae.
Entanglement can also be restored during the transmission of information (catalytic QECC)~\cite{deve,brun}.

If entanglement is assumed to be
an initial resource, QECC and QT are no longer equivalent. A physical quantum channel
is not needed at all for QT, since the quantum state has not to be sent through it.
The communication channel is rather determined by the entangled pair itself.
If a maximally entangled pair is available, this channel is given by ${\mathcal I}$;
otherwise, it will be a non-ideal operation.

These remarks clarify that the resources available to sender and receiver are fundamental to study methods
for improving the performances of noisy channels. In fact, they can be used to modify the physical channel
$\varepsilon$, and obtain a new channel with improved performance.

The aim of this work is to characterize in a general setting the impact of the available
resources on this channel, $\tilde{\varepsilon} = \tilde{\varepsilon} (\varepsilon, u)$.
This analysis shreds light on the existing techniques used to fight decoherence in noisy channels. In
particular, it describes any protocol making a single use of the physical channel, with local
operations performed instantaneously. Moreover, it represents the starting point for a systematic study
of the control and the engineering of noisy channels, since their manipulations are clearly related to the
available resources. In this paper, we present some results concerning the reliable transmission of quantum
information through channels affected by arbitrary noise, or by a specific decohering action. The latter case
is particularly relevant: to determine the best strategy (that is, the cheapest in terms of resources)
for sending quantum information through channels affected by specific models of noise is still an open problem.
In this case, the resources needed for QT or QECC are not optimal, since these protocols are aimed to correct
universal noise.

The paper is organized as follows. In Section \ref{sec2}, after introducing the notation, we develop the general formalism.
We derive the form of the operation mapping $\varepsilon$ into the new channel $\tilde{\varepsilon}$ in terms
of the control actions, that is the available resources, and study its properties. A convenient expression of this completely
positive map is obtained by using the Choi-Jamiolkowski isomorphism~\cite{jami}. In Section \ref{sec3},
we describe how to engineer a specific channel manipulation, and introduce the related notion of complexity
of a noisy channel.
In Section \ref{sec4}, we describe how QT fits into our formalism. In particular, we
discuss the case of noisy channels with maximal complexity, for which the
resources needed for QT are necessary and sufficient to deterministically obtain
$\tilde{\varepsilon} = {\mathcal I}$ starting with an arbitrary $\varepsilon$. Therefore,
QT is the only deterministic protocol that perfectly sends quantum information
through an arbitrary noisy channel without multiple use of it. Moreover,
we discuss the case of QT without maximal entanglement as a protocol for reducing the
complexity of a noisy channel, and complete the discussion in Section \ref{sec5}, where
the connection between our formalism and the theory of QECC is established. Since
all the examples of protocols presented in former sections are based on QT, in Section
\ref{sec6} we provide further examples in which the quantum state is effectively sent
through the physical channel. Finally, in Section \ref{sec8} we discuss our results, and
mention some possible generalizations of the formalism presented in this work.


\section{Noisy channels and available resources}\label{sec2}

\begin{figure}[t]
\centering
\begin{picture}(140,40)(0,10)

\linethickness{0,35pt}
\put(0,25){\line(1,0){50}}
\put(90,25){\line(1,0){50}}
\put(50,17){\framebox(40,16)[]{$\varepsilon$}}
\put(5,28){\makebox(0,0)[b]{$\rho^A$}}
\put(139,28){\makebox(0,0)[b]{$\rho^B$}}
\put(5,19){\makebox(0,0)[b]{$\rightarrow$}}
\put(135,19){\makebox(0,0)[b]{$\rightarrow$}}

\end{picture}
\caption{Noisy channel $\varepsilon$.}\label{fig1}
\end{figure}
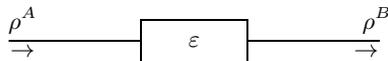

We consider a quantum channel connecting a sender $A$ and a receiver $B$,
described by a completely positive map $\varepsilon$, that can be written
using the operator sum representation~\cite{krau} as
\begin{equation}\label{krauchan}
    \rho^B = \varepsilon [\rho^A] = \sum_i E_i \rho^A E_i^{\dagger},
\end{equation}
where the Kraus operators $E_i$ satisfy $\sum_i E_i^{\dagger} E_i \leqslant I$, and equality holds for trace-preserving maps.
In the trace-decreasing case, ${\rm Tr} \rho^B$ is the probability to obtain the normalized state corresponding to
$\rho^B$. For simplicity, in the following we will mostly assume that $\varepsilon$ is trace-preserving.
The trace-class operators $\rho^A$ and $\rho^B$ are density matrices representing the state of the system before
and after the transmission through the channel respectively. These operators act on Hilbert spaces ${\mathcal H}_A$ and
${\mathcal H}_B$, and we assume that ${\rm dim} {\mathcal H}_A = {\rm dim} {\mathcal H}_B = N$.
The channel is shown in Fig. \ref{fig1}. In general, many different physical processes
could affect the transmission; the form of $\varepsilon$ could also be unknown. To
deal with these situations, we find convenient to describe a general quantum channel
as the family of completely positive maps
\begin{equation}\label{changen}
    {\mathcal E} = \{\varepsilon_i: {\mathcal T}({\mathcal H}_A) \rightarrow {\mathcal T}({\mathcal H}_B); i \in {\mathcal S}\},
\end{equation}
where ${\mathcal S}$ is an arbitrary set, and ${\mathcal T} ({\mathcal H}_x)$ denotes the set of trace-class operators
on the Hilbert space ${\mathcal H}_x$. 

We assume that $A$ and $B$ can instantaneously perform the following control operations on their systems:
\begin{enumerate}
  \item[(a)] Couple $\rho^A$ and $\rho^B$ with ancillae, represented by the state $\rho^{ab}$, where $a$ and $b$
  label the parts owned by $A$ and $B$, respectively, with Hilbert spaces ${\mathcal H}_a$ and ${\mathcal H}_b$.
  If $\rho^{ab}$ is an inseparable state, there is initial entanglement shared between sender and receiver.
  \item[(b)] Discard parts of their systems, corresponding to a partial trace operation.
  \item[(c)] Apply local unitaries $U^{Aa}$ and $U^{Bb}$ to the part they own, including ancillae.
  These operators act on the spaces ${\mathcal H}_A \otimes {\mathcal H}_a$ and ${\mathcal H}_B \otimes {\mathcal H}_b$,
  respectively.
  \item[(d)] Perform projective measurements, represented by projectors $\Pi^{Aa}$ and $\Pi^{Bb}$,
  on the part they own, and inform the other party about the measurement outcomes through noiseless classical
  communication (CC). This information can be used to correlate the operations performed by $A$ and $B$.
\end{enumerate}

The families of operations (a)-(d) are not all independent. For example, non-selective measurements can be described
as unitary operations acting on enlarged systems, followed by a partial trace operation. Analogously,
correlated operations performed by $A$ and $B$ can be obtained by using particular ancillae $\rho^{ab}$,
rather than CC. Notice that any completely positive operation and generalized measurement (POVM) can be
separately performed by $A$ and $B$. (In the literature, operations (a)-(d) are often denoted by LOCC).

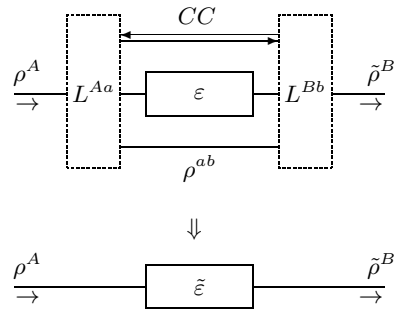
\begin{figure}[t]
\centering
\begin{picture}(140,130)(0,10)

\linethickness{0,35pt}
\put(0,25){\line(1,0){50}}
\put(90,25){\line(1,0){50}}
\put(50,17){\framebox(40,16)[]{$\tilde{\varepsilon}$}}
\put(5,28){\makebox(0,0)[b]{$\rho^A$}}
\put(139,28){\makebox(0,0)[b]{$\tilde{\rho}^B$}}
\put(5,19){\makebox(0,0)[b]{$\rightarrow$}}
\put(135,19){\makebox(0,0)[b]{$\rightarrow$}}

\linethickness{0,35pt}
\put(0,98){\line(1,0){20}}
\put(40,98){\line(1,0){10}}
\put(90,98){\line(1,0){10}}
\put(120,98){\line(1,0){20}}
\put(40,78){\line(1,0){60}}
\put(40,118){\vector(1,0){60}}
\put(100,120.5){\vector(-1,0){60}}
\put(50,90){\framebox(40,16)[]{$\varepsilon$}}
\put(5,101){\makebox(0,0)[b]{$\rho^A$}}
\put(139,101){\makebox(0,0)[b]{$\tilde{\rho}^B$}}
\put(20,70){\dashbox{1}(20,58)[]{$L^{Aa}$}}
\put(100,70){\dashbox{1}(20,58)[]{$L^{Bb}$}}
\put(70,65){\makebox(0,0)[b]{$\rho^{ab}$}}
\put(69,124.5){\makebox(0,0)[b]{$CC$}}
\put(68,43){\makebox(0,0)[b]{$\Downarrow$}}
\put(5,92){\makebox(0,0)[b]{$\rightarrow$}}
\put(135,92){\makebox(0,0)[b]{$\rightarrow$}}

\end{picture}
\caption{The noisy channel $\varepsilon$, modified using the available resources represented by the control
parameters $u$, defines the new channel $\tilde{\varepsilon} = \tilde{\varepsilon} (\varepsilon, u)$.}\label{fig2}
\end{figure}

By using the resources (a)-(d), we can define a new quantum channel $\tilde{\varepsilon}$
\begin{equation}\label{superp}
    \tilde{\rho}^B = \tilde{\varepsilon} [\rho^A] = \sum_{\eta = 1}^M \tilde{\varepsilon}_{\eta} [\rho^A],
\end{equation}
where $M$ different operations of the form
\begin{equation}\label{chan2}
    \tilde{\varepsilon}_{\eta} [\rho^A] = {\rm Tr}_{ab} \Bigl( L^{Bb}_{\eta} \varepsilon [L^{Aa}_{\eta} \rho^A \otimes
    \rho^{ab} L^{Aa \dagger}_{\eta}] L^{Bb \dagger}_{\eta} \Bigr)
\end{equation}
are superimposed, depending on the results of the projective measurement given by
$\{ \Pi_{\eta}^{Aa}, \eta\}$, and the manipulations performed by $A$ and $B$ are correlated via CC.
We use the notation $L^{Aa}_{\eta} = \Pi^{Aa}_{\eta} U^{Aa}$,
and analogously for $B$, with the difference that the unitary operation performed by $B$
may depend on $\eta$; see Fig. \ref{fig2}. The relations (\ref{superp}) and (\ref{chan2})
express the most general transformation of the noisy channel, and define a map $\varepsilon \rightarrow \tilde{\varepsilon}$.
The corresponding protocol is called {\it deterministic} if $\tilde{\varepsilon}$ is trace-preserving,
and {\it probabilistic} otherwise. In this case, the probability of success is given by ${\rm Tr \tilde{\rho}^B}$.

While any local operation can be written as a successive application of a unitary operator and a projection~\cite{niel},
these two operations do not necessarily represent the real physical manipulations the system has been
subject to. In some cases it can be useful to write the local operations in a different form, maybe
introducing additional indices into the formalism. In view of practical realizations, it
is often preferable to write the local operation in terms of a redefined unitary operation followed by a projection
involving only the ancilla, and additional post-selection or not, depending on the specific situation at
hand. In fact, this implementation do not ask for a non-destructive joint $\Pi^{Aa}$.

In order to simplify the notation, in the following we assume that $M = 1$, and drop the index $\eta$.
The general case is recovered by restoring this index and the corresponding sum.

For the auxiliary systems, we consider an arbitrary pure state
$\rho^{ab} = \vert \psi \rangle^{ab} \langle \psi \vert$. The generalization to mixed states
follows from linearity of (\ref{chan2}).
Without loss of generality we can write
\begin{equation}\label{schmidt}
    \vert \psi \rangle^{ab} = \sum_{k = 0}^{N - 1} \mu_k \vert k \rangle^a \otimes \vert k \rangle^b,
\end{equation}
where $\mu_k$, $k = 0, \ldots, N - 1$ are the real Schmidt coefficients.
$\vert \psi \rangle^{ab}$ is maximally
entangled if and only if $\mu_k = \frac{1}{\sqrt{N}}$ for all $k$; it is a
product state if and only if $\mu_k = \delta_{k \bar{k}}$ for some
$\bar{k} \in \{0, \ldots, N - 1\}$. These parameters will quantify the impact of the initially shared
entanglement between $A$ and $B$ on the control of the noisy channel. We then rewrite (\ref{superp}) as
\begin{equation}\label{chan3}
    \tilde{\rho}^B = \sum_{i,j} \mu_i \mu_j \sum_{k,l} B_{k,i} \varepsilon [A_{l,i}
    \rho^A A^{\dagger}_{l,j}] B^{\dagger}_{k,j},
\end{equation}
where we have defined the operators
\begin{equation}\label{auxop}
    A_{i,j} = \langle i \vert^a L^{Aa} \vert j \rangle^a, \quad B_{i,j} = \langle i \vert^b L^{Bb} \vert j \rangle^b
\end{equation}
acting on the Hilbert spaces ${\mathcal H}_A$ and ${\mathcal H}_B$ respectively. These operators
satisfy
\begin{equation}\label{auxop2}
    A_{i,j} = \sum_k \alpha_{i,k} a_{k,j}, \quad B_{i,j} = \sum_k \beta_{i,k} b_{k,j},
\end{equation}
where
\begin{equation}\label{auxop3}
    \alpha_{i,j} = \langle i \vert^a \Pi^{Aa} \vert j \rangle^a, \quad \beta_{i,j} = \langle i \vert^b \Pi^{Bb} \vert j \rangle^b
\end{equation}
are related to the non trace-preserving part of the local operation, whereas
\begin{equation}\label{auxop4}
    a_{i,j} = \langle i \vert^a U^{Aa} \vert j \rangle^a, \quad b_{i,j} = \langle i \vert^b U^{Bb} \vert j \rangle^b
\end{equation}
depend on its unitary part. Since $\Pi^{Aa \dagger} = \Pi^{Aa}$, and $(\Pi^{Aa})^2 = \Pi^{Aa}$, we have that
\begin{equation}\label{relpro}
    \alpha_{i,j} = \alpha_{j,i}^{\dagger}, \quad \alpha_{i,j} = \sum_k \alpha_{i,k} \alpha_{k,j},
\end{equation}
and analogous formulas hold for $\beta_{i,j}$.
Moreover, $U^{Aa \dagger} U^{Aa} = U^{Aa} U^{Aa \dagger} = I$ implies that
\begin{equation}\label{reluni}
    \sum_k a_{i,k} a_{j,k}^{\dagger} = \sum_k a_{k,i}^{\dagger} a_{k,j} = \delta_{ij} I^A,
\end{equation}
and similar relations for $b_{i,j}$.
For further reference, we observe that the unitarity of $U^{Aa}$ implies the unitarity of
the adjoint (or the transpose) of $U^{Aa}$ with respect to the degrees of freedom of $A$ (or $a$).
It follows that
\begin{equation}\label{reluni2}
    \sum_k a_{i,k}^{\dagger} a_{j,k} = \sum_k a_{k,i} a_{k,j}^{\dagger} = \delta_{ij} I^A,
\end{equation}
and analogously for $b_{i,j}$.
The operator sum representation of the modified channel is then given by
\begin{equation}\label{krauchan2}
    \tilde{\rho}^B = \tilde{\varepsilon} [\rho^A] = \sum_{i,k,l} \tilde{E}_{i,k,l} \rho^A \tilde{E}_{i,k,l}^{\dagger},
\end{equation}
with Kraus operators
\begin{equation}\label{krauop2}
    \tilde{E}_{i,k,l} = \sum_{j} \mu_j B_{k,j} E_i A_{l,j}.
\end{equation}
The new channel $\tilde{\varepsilon}$ is trace-preserving (that is, the control protocol is deterministic)
if $\varepsilon$ and the local operations performed by $A$ and $B$ are trace-preserving, that is $\Pi^{Aa}
= I^{Aa}$ and $\Pi^{Bb} = I^{Bb}$. For $M > 1$, a trace-preserving modified channel can be obtained
by using $\Pi^{Bb}_{\eta} = I^{Bb}$, and $\sum_{\eta} \Pi^{Aa}_{\eta} =
I^{Aa}$, leading to
\begin{equation}\label{relproeta}
    \sum_{\eta} \sum_k A_{i,k}^{\eta} A_{j,k}^{\eta \dagger} = \delta_{ij} I^A.
\end{equation}
This scenario correspond to one-way CC from sender to receiver, used to
correlate the operations of $A$ and $B$, when to every outcome in $A$ is
associated a trace-preserving operation in $B$.

If $\rho^{ab}$ is a product state, that is $\mu_k = \delta_{k \bar{k}}$, the evolution takes the form
\begin{equation}\label{pure}
    \tilde{\rho}^B = \sum_{k,l} B_{k,\bar{k}} \varepsilon [A_{l,\bar{k}}
    \rho^A A^{\dagger}_{l,\bar{k}}] B^{\dagger}_{k,\bar{k}}.
\end{equation}
In this case we can write $\tilde{\varepsilon} = \varepsilon^B \circ \varepsilon \circ \varepsilon^A$, where $\varepsilon^A$
and $\varepsilon^B$ have Kraus operators $A_{i,\bar{k}}$ and $B_{i,\bar{k}}$, respectively. The original noisy channel is modified
by the application of control operations before and after it. We observe that an arbitrary separable
state, $\rho^{ab} = \sum_i p_i \rho_i^a \otimes \rho_i^b$, generates
\begin{equation}\label{clacor}
    \tilde{\varepsilon} = \sum_i p_i \varepsilon^B_i \circ \varepsilon \circ \varepsilon^A_i,
\end{equation}
and the trace-preserving operations applied before and after the noisy channel,
$\varepsilon_i^A$ and $\varepsilon_i^B$, are classically correlated. Viceversa,
if $\tilde{\varepsilon}$ can be expressed as in (\ref{clacor}), with $\varepsilon_i^A$ and
$\varepsilon_i^B$ local, trace-preserving operations, then there exist a separable state $\rho^{ab}$ and
local unitary operations $L^{Aa}$ and $L^{Bb}$ such that $\tilde{\varepsilon}$ can be written as
in (\ref{superp}), with $M = 1$. We conclude that, in order
to create non-classical correlations between their operations, $A$ and $B$ must use entangled ancillae.

To study the properties of the map $\varepsilon \rightarrow \tilde{\varepsilon} = \tilde{\varepsilon} (\varepsilon, u)$,
we find convenient to use the Choi-Jamiolkowski isomorphism~\cite{jami,arri,dur} between completely positive maps
$\varepsilon: {\mathcal H}_A \rightarrow {\mathcal H}_B$ and positive operators $R^{BA}$ acting on ${\mathcal H}_B
\otimes {\mathcal H}_A$, defined by
\begin{equation}\label{cjiso}
    R^{BA} = \varepsilon \otimes I^A [\Psi_0^{AA}],
\end{equation}
where ${\Psi}_0^{xy} = \vert \psi_0 \rangle^{xy} \langle \psi_0 \vert$ is a maximally entangled state
for the systems $x$ and $y$ (in (\ref{cjiso}) two copies of the system $A$ are considered), such that
\begin{equation}\label{bell0}
    \vert \psi_0 \rangle^{xy} = \frac{1}{\sqrt{N}} \sum_{i = 0}^{N - 1} \vert i \rangle^x \otimes \vert i \rangle^y.
\end{equation}
By using this isomorphism, equation (\ref{superp}) can be conveniently expressed
by a completely positive map $\lambda$ such that
\begin{equation}\label{jammod}
    \tilde{R}^{BA} = \lambda[R^{BA}] = \sum_{\eta} \sum_{k,l} \Lambda^{\eta}_{k,l} R^{BA} \Lambda^{\eta \dagger}_{k,l}
\end{equation}
where we assumed that $R^{BA} \simeq \varepsilon$ and $\tilde{R}^{BA} \simeq \tilde{\varepsilon}$,
and introduced the Kraus operators
\begin{equation}\label{kraujan}
    \Lambda_{k,l}^{\eta} = \sum_i \mu_i B^{\eta}_{k,i} \otimes A^{\eta T}_{l,i}.
\end{equation}
In general, the map $\lambda$ is not trace-preserving, and it can be trace-increasing. In fact,
by choosing $\mu_k = \delta_{k \bar{k}}$, it is possible to prove the upper bound
\begin{equation}\label{trajan2}
    \sum_{\eta} \sum_{k,l} \Lambda_{k,l}^{\eta \dagger} \Lambda_{k,l}^{\eta} \leqslant ({\rm dim} {\mathcal H}_a) \, I^{BA}.
\end{equation}
If we choose ${\rm dim}{\mathcal H}_A = {\rm dim}{\mathcal H}_a$ and the local operators
\begin{equation}\label{loop}
    U^{Aa} = U^{Aa}_{swap}, \qquad
    \Pi_{\eta}^{Aa} = \vert \eta \rangle^A \langle \eta \vert \otimes I^a,
\end{equation}
we obtain
\begin{equation}\label{trajan}
    \sum_{\eta} \sum_{k,l} \Lambda^{\eta \dagger}_{k,l} \Lambda^{\eta}_{k,l} = ({\rm dim}{\mathcal H}_a) \, I^B \otimes
    \Bigl( \sum_i \mu_i^2 \vert i \rangle^A \langle i \vert \Bigr),
\end{equation}
leading to at least one eigenvalue exceeding $1$ unless the probe is in a maximally entangled state. The swap operator
is defined by
\begin{equation}\label{swap}
    U_{swap}^{xy} \vert \varphi_1 \rangle^x \otimes \vert \varphi_2 \rangle^y =
    \vert \varphi_2 \rangle^x \otimes \vert \varphi_1 \rangle^y.
\end{equation}

We provide now some examples of trace-preserving maps $\lambda$. We begin by assuming
that $M = 1$, and $\Pi^{Aa} = \Pi^{Bb} = I$, that is, only
trace-preserving operations can be performed by sender and receiver. In this case,
$A_{i,j} = a_{i,j}$, $B_{i,j} = b_{i,j}$, and the formulas characterizing the
modified channel $\tilde{\varepsilon}$ highly simplify. In particular, the
Kraus operators (\ref{krauop2}) satisfy
\begin{equation}\label{sumkrau2un}
    \sum_{i,k,l} \tilde{E}_{i,k,l}^{\dagger} \tilde{E}_{i,k,l} = \sum_j \mu_j^2 \sum_{l} a_{l,j}^{\dagger}
    \Bigl( \sum_i E_i^{\dagger} E_i \Bigr) a_{l,j},
\end{equation}
and $\tilde{\varepsilon}$ is a trace-preserving channel
if and only if $\varepsilon$ is trace-preserving. The dependence on the controls
acting before the noisy channel, and on the entanglement resource,
is relevant only for non trace-preserving channels $\varepsilon$.
Correspondingly, equation (\ref{trajan}) becomes
\begin{equation}\label{trajanun}
    \sum_{k,l} \Lambda_{k,l}^{\dagger} \Lambda_{k,l} = I^{BA},
\end{equation}
and $\lambda$ is a trace-preserving completely positive map.

A similar situation arises also when $M > 1$, if we consider the scenario of QT,
that is $U^{Aa}_{\eta} = I^{Aa}$ and $\Pi_{\eta}^{Aa}$ are the projectors on the
Bell basis. An explicit computation proves that
\begin{equation}\label{trajanun2}
    \sum_{\eta} \sum_{k,l} \Lambda_{k,l}^{\eta \dagger} \Lambda_{k,l}^{\eta} = I^{BA}.
\end{equation}
More generally, a maximally entangled probe is a sufficient condition for trace-preservation,
as long as $\sum_{\eta} \Pi_{\eta}^{Aa} = I^{Aa}$.

Another sufficient condition for a trace-preserving $\lambda$ is that the
operation performed by $A$ is bistochastic. In our notation, this property reads
\begin{equation}\label{relproeta2}
    \sum_{\eta} \sum_k A_{i,k}^{\eta \dagger} A_{j,k}^{\eta} = \delta_{ij} I^A,
\end{equation}
and (\ref{trajanun2}) follows, independent of the entanglement shared between $A$ and $B$.

Notice that there is not a simple relation between the property  of trace-preservation for $\lambda$ and
the same property for the channels. QT represents a good example, since
the corresponding $\lambda$ is trace-preserving, and it can map non trace-preserving
channels $\varepsilon$ to the trace-preserving ideal operation $\tilde{\varepsilon} =
{\mathcal I}$ by completely replacing $\varepsilon$ by LOCC (the formalism of QT is reviewed in Section \ref{sec4}).

In summary, the problem of improving the performance of a quantum channel $\varepsilon$ by using
a set of available resources is equivalent to the problem of steering in a controlled way the
corresponding Choi-Jamiolkowski state $R^{BA}$ to $\tilde{R}^{BA} = \lambda [R^{BA}]$.
Following (\ref{kraujan}), the resources accessible to $A$ and $B$ determine the form of the Kraus
operators $\Lambda_{k,l}$ that describe the completely positive transformation (\ref{jammod})
between the initial and the controlled Choi-Jamiolkowski state.

It is now interesting to study how it is possible to control the properties and improve the
performance of a noisy channel $\varepsilon$. In principle several quantities
that characterize the channel could be used as a figure of merit, for example
its capacity, an operator distance between $\varepsilon$
and $\mathcal I$, the average fidelity of transmission, etc. Following the
ideas developed in this section, it is natural to look for quantities inspired by the
Choi-Jamiolkowski isomorphism. For instance, one could consider the Choi-Jamiolkowski
fidelity between $\tilde{R}^{BA} (u)$ and $\Psi_0^{BA}$, given by
\begin{equation}\label{cjfid}
    {\mathcal F} (u) = {\rm Tr} \Bigl( \tilde{R}^{BA} (u) \Psi_0^{BA} \Bigr) = \langle \psi_0 \vert^{BA}
    \tilde{R}^{BA} (u) \vert \psi_0 \rangle^{BA},
\end{equation}
and optimize it with respect to the control parameters $u$ for a specific scenario
described by $\varepsilon$ and given resources. This would enable us to examine how close to the
identity channel an initial channel can be made, thus describing a potential noise reduction
through our control resources. However, in this work, we choose to
introduce a new figure of merit, connected to the structure of the noise rather that to its
size. Later on, we will describe the meaning of this quantity in connection with
geometric control and QECCs.


\section{Complexity of a noisy channel}\label{sec3}

In this section, as a reference quantity, we define the notion of {\it complexity} of the
noisy channel, strictly related to its Kraus rank, and representing, in some sense,
the pattern of the channel. In defining this quantity, we want to take into account
the fact that the map $\varepsilon$ associated to the noisy channel could be unknown.
In this case, following (\ref{changen}), a family of maps $\mathcal E$ has to be taken into account.
The motivation for adopting this quantity as a figure of merit is that in this work we consider a
geometric control perspective~\cite{dale}, and the complexity will provide a quantitative measure of the
difficulty in engineering a specific map $\lambda$ through the available resources
$L^{Aa}$, $L^{Bb}$, and $\rho^{ab}$.

To start with, we assume that $\varepsilon$ is known, and denote by $R^{BA}$ the
corresponding Choi-Jamiolkowski state. Its spectral decomposition is given by
\begin{equation}\label{spedecj}
    R^{BA} = \sum_j r_j \vert r_j \rangle^{BA} \langle r_j \vert,
\end{equation}
where $\{\vert r_j \rangle^{BA}; j = 0, \ldots, N^2 - 1\}$ is an orthonormal basis in ${\mathcal H}_A \otimes
{\mathcal H}_B$, $0\leqslant r_j \leqslant 1$, and $\sum_j r_j = 1$, since we assumed that $\varepsilon$ is trace-preserving.

The optimal manipulation of a noisy channel would map it to the ideal channel, $\varepsilon \rightarrow \tilde{\varepsilon} =
{\mathcal I}$, whose Choi-Jamiolkowski state is given by the Bell state ${\Psi}_0^{BA} = \vert \psi_0
\rangle^{BA} \langle \psi_0 \vert$. Therefore, in the case without CC ($M = 1$), we require
\begin{equation}\label{aim}
    \Psi_0^{BA} = \lambda [R^{BA}] = \sum_j \Lambda_j R^{BA} \Lambda_j^{\dagger},
\end{equation}
that is,
\begin{equation}\label{aim2}
    \Lambda_j \vert r_i \rangle^{BA} = \alpha_j^i \vert \psi_0 \rangle^{BA}
\end{equation}
for all $i$ with $r_i \ne 0$, since we are only interested in the action of $\lambda$ on
$R^{BA}$, and $\sum_j \vert \alpha_j^i \vert^2 \leqslant 1$. In particular, equality holds
for deterministic protocols; for probabilistic protocols, $\sum_j \vert \alpha_j^i \vert^2$
represents the probability of success. In this section we will always consider deterministic protocols. By using
the invariance of $\lambda$ under the transformation $\Lambda_j \rightarrow \sum_j u_{ij} \Lambda_j$,
where the coefficients $u_{ij}$ form a unitary matrix, and limiting our attention to trace-preserving maps,
it is possible to rewrite (\ref{aim2}) as
\begin{equation}\label{aim3}
    \Lambda_j \vert r_i \rangle^{BA} = \delta_{ij} \vert \psi_0 \rangle^{BA}.
\end{equation}
In the case of $M > 1$, this condition generalizes to
\begin{equation}\label{aim3bis}
    \Lambda_j^{\eta} \vert r_i \rangle^{BA} = \delta_{ij} \beta_{\eta} \vert \psi_0 \rangle^{BA},
\end{equation}
with $\eta = 1, \ldots, M$, and $\sum_{\eta} \vert \beta_{\eta} \vert^2 = 1$.
Therefore, the rank of $R^{BA}$ (also called the Kraus rank of $\varepsilon$) is a lower bound for the
number of Kraus operators appearing in the map $\lambda$. Therefore, this map can be unitary only if $R^{BA}$ is a pure state.
In general, there are several maps $\lambda$ satisfying (\ref{aim}), and they can
be trace-preserving or not. Correspondingly, there is a richer set of resources that
can be used in order to perform the desired manipulation of
the channel. This freedom decreases as the rank of $R^{BA}$
increases. If the rank of $R^{BA}$ is maximal, there is only one map $\lambda$ consistent
with the constraints (\ref{aim3}). A more rigorous analysis is presented in Appendix \ref{app1},
where we derive the upper bound for the number of Kraus operators $\Lambda_j$ that can appear in $\lambda$.

This discussion motivates the following definition of {\it complexity} of a noisy channel ${\mathcal E}$.

\begin{definition}
Given a noisy channel described by a family of completely positive maps
${\mathcal E} = \{\varepsilon_i; i \in {\mathcal S}\}$,
with corresponding Choi-Jamiolkowski states $R^{BA}_i \simeq \varepsilon_i$, its complexity $\chi ({\mathcal E})$
is given by
\begin{equation}\label{defcomp}
    \chi ({\mathcal E}) = {\rm dim} \bigoplus_{i \in {\mathcal S}} {\mathcal V}_i,
\end{equation}
where ${\mathcal V}_i \subseteq {\mathcal H}_B \otimes {\mathcal H}_A$ is the support of $R^{BA}_i$,
and $i \in {\mathcal S}$. If ${\mathcal V}_i = {\rm span} \{ \vert \psi_0 \rangle^{BA} \}$ for all $i$,
we define $ \chi ({\mathcal E}) = 0$. If the set ${\mathcal E}$ contains only one element
$\varepsilon$, we will use the notation $\chi (\varepsilon)$.
\end{definition}

The complexity of a noisy channel $\varepsilon$ closely resembles its Kraus rank, that is,
the dimension of the support of the corresponding Choi-Jamiolkowski state $R^{BA}$.
An exception is given by $R^{BA} = \Psi_0^{BA}$, where $\lambda = I^{BA}$ is obtained
with trivial controls, that is without any ancilla $\rho^{ab}$, and local operations $L^A = I^A$,
and $L^B = I^B$. In this case, the complexity vanishes by definition, but the Kraus rank does not.

The notion of complexity is meaningful even when the action of the noisy channel is not known,
or different maps $\varepsilon_i$ can act during the transmission of quantum information. In this case,
condition (\ref{aim}) has to be imposed for all these possible processes. The general definition of
complexity automatically takes into account the minimal subspace of the Hilbert space, including all the
supports of the states $R^{BA}_i$, corresponding to $\varepsilon_i$.

In general, $0 \leqslant \chi (\varepsilon) \leqslant N^2$. In particular, $\chi (\varepsilon) = 0$ if an
only if $R^{BA} = \Psi_0^{BA}$, $\chi (\varepsilon) = 1$ if an only if $R^{BA}$ is
a pure state and $R^{BA} \ne \Psi_0^{BA}$, and $\chi (\varepsilon) =  N^2$ if and only if the support of $R^{BA}$
is the whole space ${\mathcal H}_B \otimes {\mathcal H}_A$. Similarly, the complexity is maximal whenever
the channel is affected by an unknown error.

In the qubit case ($N = 2$), the maximal complexity of a noisy channel is $\chi (\varepsilon) = 4$.
This value characterizes, for example, the depolarizing channel. Other standard channels are
the bit-flip, phase-flip, amplitude-damping, and phase-damping channels; all of them have $\chi (\varepsilon) = 2$.
If a channel can be affected by depolarization or phase-flip errors, its complexity will be $\chi (\mathcal E) = 4$.
The same value characterizes a channel subject to bit-flip or amplitude-damping. If bit-flip and
phase-flip errors are acting, $\chi (\mathcal E) = 3$. Notice that a larger value of the complexity
does not necessarily mean that the fidelity of transmission, or the capacity of the channel, are
smaller. In fact, the complexity does not incorporate the size of the decohering effect
of the environment, {\it but only the structure of the noise it induces}. A similar situation is encountered
in QECC, where a 3-qubit code can perfectly correct a single, strong bit-flip error, but does not protect against
a weak depolarization errors, not even in a single channel.

Notice that, if the resources are limited, it could be possible
to reduce the complexity of a channel $\varepsilon$, but never get $\chi (\tilde{\varepsilon}) = 0$.
However, as long as $\chi (\tilde{\varepsilon}) < \chi (\varepsilon)$, the control on the channel
simplifies its structure and reduces the needed resources for the corresponding QECC.
An example is provided in Section \ref{sec5}.
Moreover, if a protocol decreases the complexity of a channel $\varepsilon$, it is in general
false that it decreases the complexity of any channel $\varepsilon^{\prime}$ with
$\chi (\varepsilon^{\prime}) < \chi (\varepsilon)$. This follows from the definition of
complexity. However, if a protocol can deterministically correct errors from a channel with maximal
complexity, then it can correct every error, that is, it is universal.


\section{Channels with maximal complexity and QT}\label{sec4}

In this section, we establish a connection between channels with maximal complexity and the
QT protocol. Given a generic channel ${\mathcal E}$ with maximal
complexity, $\chi (\mathcal{E}) = N^2$, what are the resources needed to
deterministically reduce its complexity? If a protocol exists,
such that the resulting complexity vanishes, it is universal, and
it can correct arbitrary errors.

An important example of a universal scheme is given by QT which we briefly outline in its standard form.
Without loss of generality, we consider a pure state $\rho^A = \vert \phi \rangle^A \langle \phi
\vert$ to be sent to $B$. The two parties share an ancilla given by the maximally entangled state $\rho^{ab} = \Psi_0^{ab}$.
Party $A$ performs a measurement on the maximally entangled basis in ${\mathcal H}_A \otimes {\mathcal H}_a$,
defined by the set of projectors $\Pi^{Aa}_{\eta} = \Psi_{\eta}^{Aa} = \vert \psi_{\eta} \rangle^{Aa} \langle \psi_{\eta} \vert$, where
\begin{equation}\label{bell}
    \vert \psi_{\eta} \rangle^{xy} = \frac{1}{\sqrt{N}} \sum_{k = 0}^{N - 1} e^{2 \pi i k \frac{n}{N}}
    \vert k \rangle^x \otimes \vert (k + m)_N \rangle^y
\end{equation}
is the generalized Bell basis for the systems $x$ and $y$, with $\eta = 0, \ldots, N^2 - 1$, $n = \eta \,{\rm div}\, N$,
$m = \eta \,{\rm mod}\, N$ (that is, $\eta = n N + m$), and we used the concise notation $(i)_N = i \,{\rm mod}\, N$.
Then, $A$ sends the measurement outcome $\eta$ to $B$ through a noiseless classical channel.
According to this result, $B$ applies the unitary operator given by
\begin{equation}\label{unitqt}
    U_{\eta}^{Bb} = U^{Bb}_{swap} \Bigl( I^B \otimes \sum_{k = 0}^{N - 1} e^{2 \pi i k \frac{n}{N}} \vert k \rangle^b \langle (k + m)_N \vert \Bigr),
\end{equation}
that reproduces the unknown initial state $\rho^A$ in his own space ${\mathcal H}_b$, and transfers
it into ${\mathcal H}_B$. The SWAP operator has been introduced for convenience of notation.
The complete procedure is described by
\begin{equation}\label{telepst}
    \tilde{\rho}^B = \tilde{\varepsilon} [\rho^A] = \sum_{\eta = 0}^{N^2 - 1} {\rm Tr}_{ab} \Bigl( U^{Bb}_{\eta} \varepsilon [ \Pi^{Aa}_{\eta} \rho^A
    \otimes \Psi_0^{ab} \Pi^{Aa}_{\eta}] U_{\eta}^{Bb \dagger} \Bigr),
\end{equation}
and $\tilde{\rho}^B$ is a perfect copy of the initial state $\rho^A$, for all noisy channels $\varepsilon$.

The resources needed for QT are then a maximally entangled pair shared between sender and receiver, the ability to send
$N$ bits of classical information, the ability to perform Bell measurements in $A$, and to apply specific operators
in $B$. Among these, the most demanding is certainly the existence of a maximally entangled pair. We now ask whether it
is possible to reduce these resources while maintaining an universal protocol. It turns out that there is not
other universal protocol, making use of a different set of resources. This result can be summarized in the following
Theorem ~\cite{roma}.

\begin{theorem}
Given a noisy channel ${\mathcal E}$ with maximal complexity, $\chi ({\mathcal E}) = N^2$,
the resources of QT are necessary and sufficient to deterministically map it to a new
channel $\tilde{\mathcal E}$ with $\chi (\tilde{\mathcal E}) = 0$.
\end{theorem}

Notice that, in our approach, only a single use of the channel is admitted. We
do not consider the possibility of encoding the quantum state to be sent into a
larger space. Therefore there is no contradiction with the fact that QECC can
provide universal protection against errors affecting the relevant system.

The most fragile resource needed in QT is the maximally entangled pair $\rho^{ab}$.
It is therefore of interest to study how the transformed channel $\tilde{\varepsilon}$
is affected by imperfect entanglement, when the QT protocol is employed. This analysis
is performed by replacing in (\ref{telepst}) the maximally entangled state $\vert \psi_0 \rangle^{ab}$
with the arbitrary pure state given in (\ref{schmidt}). We obtain the channel
\begin{equation}\label{qtrevfinkra}
    \tilde{\rho}^B = \tilde{\varepsilon} [\rho^A] = \sum_{\eta = 0}^{N^2 - 1} \tilde{E}_{\eta} \rho^A \tilde{E}_{\eta}^{\dagger},
\end{equation}
where the action of the Kraus operators on the basis of ${\mathcal H}_A$, defined by the Bell measurement, is given by
\begin{equation}\label{kraqt}
    \tilde{E}_{\eta} \vert l \rangle^A = \frac{1}{\sqrt{N}} \mu_{(l + \eta)_N} {\mathcal I} \, \vert l \rangle^A.
\end{equation}
Since $\tilde{E}_{(\eta + N)_{N^2}} = \tilde{E}_{\eta}$,
we can simplify these expressions to
\begin{equation}\label{qtrevfinkra2}
    \tilde{\varepsilon} [\rho^A] = \sum_{j = 0}^{N - 1} \tilde{E}_j \rho^A \tilde{E}_j^{\dagger},
\end{equation}
and
\begin{equation}\label{kraqt2}
    \tilde{E}_j \vert l \rangle^A = \mu_{(l + j)_N} {\mathcal I} \, \vert l \rangle^A.
\end{equation}
Notice that when the ancilla is the maximally entangled state with $\mu_j = \frac{1}{\sqrt{N}}$ for all $j$,
$\tilde{\varepsilon}$ reduces to ${\mathcal I}$, as expected. Conversely, when there is no entanglement,
$\tilde{\varepsilon}$ acts as
\begin{equation}\label{depocha}
    \rho^A = \sum_{ij} \rho_{ij} \vert i \rangle^A \langle j \vert \rightarrow
    \tilde{\rho}^B = \sum_i \rho_{ii} \vert i \rangle^B \langle i \vert.
\end{equation}
This channel reliably transmits only classical information (the diagonal entries of $\rho^A$ in the
basis specified by the Bell measurement), as there are no quantum correlations
in $\rho^{ab}$. Finally, if the shared entangled state is mixed, the channel corresponding to QT
is a depolarizing channel, with $\tilde{\rho}^B = \frac{1}{N} I^B$ in the special case of maximal mixing.

Following (\ref{qtrevfinkra2}), for a pure $\rho^{ab}$, the new channel has complexity
$\chi (\tilde{\varepsilon}) \leqslant N$ regardless of the entanglement shared between $A$ and $B$.
Therefore, the complexity of any physical channel with $\chi (\varepsilon) > N$ can be deterministically
reduced by using the resources of QT, but imperfect (or even vanishing) entanglement.
In this process, an actual physical channel is not used at all, as well known in QT.
Reduction of complexity does not imply, in general, more reliable transmission of quantum information
(that is, transmission with higher fidelity). It rather means that further manipulations of
the channel may be performed more easily. Therefore, it depends on the specific situation whether
it is better to use the physical channel directly or the channel defined by the QT protocol.

In the case of a non-maximally entangled $\rho^{ab}$, the protocol of conclusive QT has
been introduced to faithfully send arbitrary states through the channel, with a success
probability less than unity~\cite{mor}. Conclusive QT is therefore an example of a probabilistic protocol
with $\chi (\tilde{\varepsilon}) = 0$ whenever it succeeds.


\section{Reduction of complexity and QECC}\label{sec5}

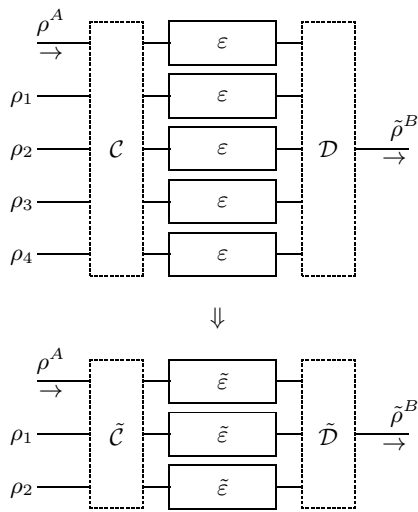
\begin{figure}[b]
\centering
\begin{picture}(140,190)(-5,-5)


\linethickness{0,35pt}

\put(0,178){\line(1,0){20}}
\put(40,178){\line(1,0){10}}
\put(90,178){\line(1,0){10}}
\put(50,170){\framebox(40,16)[]{$\varepsilon$}}

\put(0,158){\line(1,0){20}}
\put(40,158){\line(1,0){10}}
\put(90,158){\line(1,0){10}}
\put(50,150){\framebox(40,16)[]{$\varepsilon$}}

\put(0,138){\line(1,0){20}}
\put(40,138){\line(1,0){10}}
\put(90,138){\line(1,0){10}}
\put(120,138){\line(1,0){20}}
\put(50,130){\framebox(40,16)[]{$\varepsilon$}}

\put(0,118){\line(1,0){20}}
\put(40,118){\line(1,0){10}}
\put(90,118){\line(1,0){10}}
\put(50,110){\framebox(40,16)[]{$\varepsilon$}}

\put(0,98){\line(1,0){20}}
\put(40,98){\line(1,0){10}}
\put(90,98){\line(1,0){10}}
\put(50,90){\framebox(40,16)[]{$\varepsilon$}}

\put(5,181){\makebox(0,0)[b]{$\rho^A$}}
\put(139,141){\makebox(0,0)[b]{$\tilde{\rho}^B$}}
\put(20,90){\dashbox{1}(20,96)[]{${\mathcal C}$}}
\put(100,90){\dashbox{1}(20,96)[]{${\mathcal D}$}}
\put(5,172){\makebox(0,0)[b]{$\rightarrow$}}
\put(135,132){\makebox(0,0)[b]{$\rightarrow$}}

\put(-6,154){\makebox(0,0)[b]{$\rho_1$}}
\put(-6,134){\makebox(0,0)[b]{$\rho_2$}}
\put(-6,114){\makebox(0,0)[b]{$\rho_3$}}
\put(-6,94){\makebox(0,0)[b]{$\rho_4$}}

\put(68,70){\makebox(0,0)[b]{$\Downarrow$}}

\put(0,10){\line(1,0){20}}
\put(40,10){\line(1,0){10}}
\put(90,10){\line(1,0){10}}
\put(50,02){\framebox(40,16)[]{$\tilde{\varepsilon}$}}

\put(0,30){\line(1,0){20}}
\put(40,30){\line(1,0){10}}
\put(90,30){\line(1,0){10}}
\put(120,30){\line(1,0){20}}
\put(50,22){\framebox(40,16)[]{$\tilde{\varepsilon}$}}

\put(0,50){\line(1,0){20}}
\put(40,50){\line(1,0){10}}
\put(90,50){\line(1,0){10}}
\put(50,42){\framebox(40,16)[]{$\tilde{\varepsilon}$}}

\put(5,53){\makebox(0,0)[b]{$\rho^A$}}
\put(139,33){\makebox(0,0)[b]{$\tilde{\rho}^B$}}
\put(20,02){\dashbox{1}(20,56)[]{$\tilde{\mathcal C}$}}
\put(100,02){\dashbox{1}(20,56)[]{$\tilde{\mathcal D}$}}
\put(5,44){\makebox(0,0)[b]{$\rightarrow$}}
\put(135,24){\makebox(0,0)[b]{$\rightarrow$}}

\put(-6,26){\makebox(0,0)[b]{$\rho_1$}}
\put(-6,06){\makebox(0,0)[b]{$\rho_2$}}

\end{picture}
\caption{In the context of QECC, a channel $\tilde{\varepsilon}$ with smaller complexity can be dealt
with by using a reduced set of resources with respect to the original channel $\varepsilon$. For example,
a $5$-qubit code is replaced by a $3$-qubit code, after the manipulation of $\varepsilon$. This can be obtained
by using QT without perfect entanglement.}\label{fig3}
\end{figure}

In general, for a given set of resources, it may be impossible to obtain a vanishing
complexity for the redefined channel $\tilde{\varepsilon}$. In this case, a deterministic
correction of the channel is not possible, and further processing of the signal is needed
to improve the reliability of the transmission. In particular, it is necessary to
send redundant information, for example, by using many times the noisy channel.

This scenario is described by the theory of QECC: the state to be sent is encoded through a
coding operation ${\mathcal C}$ in a larger Hilbert space, whose elementary constituents are
sent through independent noisy channels. A decoding and recovering operation
${\mathcal D}$ on the receiver side completes the scheme. The global procedure can
deterministically correct the noisy channel when the error acts only on a limited number of
channels involved in the transmission, or improve the fidelity of the transmission, when only a
small error acts in every channel. Notice that QECC cannot be directly incorporated into
our formalism which considers only a single use of the channel.

A fundamental property of QECCs is the following. If a code is able to correct a particular set of
error generators (Kraus operators), every linear superposition of elements of this set
is also a correctable error for that code. It follows that, by definition, the complexity of a noisy
channel is directly related to the structure of the QECC that can correct it. Therefore, by mapping
the channel $\varepsilon$ into $\tilde{\varepsilon}$ with reduced complexity, in principle, a
simpler code is needed, that is a code using a reduced number of ancillae.

We provide a simple example of this procedure, by considering the transmission of a qubit $\rho^A$ ($N = 2$) through
an arbitrary noisy channel $\varepsilon$, with complexity $\chi (\varepsilon) = 4$.
In the standard setting, the minimal code that has to be used requires $4$ additional qubits
to define the code space. We assume now that sender and receiver share some pure, non-maximally
entangled pairs and use them to perform QT on the states they want to send. Following Section \ref{sec5},
this procedure defines a phase-flip channel for each entangled pair, that is
\begin{equation}\label{depha2}
    \tilde{\varepsilon} [\rho] = (1 - p_{\mu}) \rho + p_{\mu} \sigma_z \rho \sigma_z,
\end{equation}
where the probability $p_{\mu}$ depends on the entanglement through the Schmidt coefficient $\mu$ as
\begin{equation}\label{pimu}
    p_{\mu} = \frac{1}{2} - \mu \sqrt{1 - \mu^2}.
\end{equation}
These channels can be used to send $\rho^A$ after encoding it in a $3$-qubit space, as only $2$ additional
ancillae are required to correct the phase-flip channel. Therefore, the number of auxiliary
systems needed to define the code space can be reduced from $4$ to $2$, and the corresponding
coding and decoding operations, $\tilde{\mathcal C}$ and $\tilde{\mathcal D}$, are simpler, see Fig. \ref{fig3}.
This is possible because of the additional resource of pure, partial entanglement. Interestingly, with the
same resource of entanglement, an analogous reduction of ancillae is obtained via Entanglement Assisted QECC~\cite{brun}.

Notice that, in this example, $\tilde{\varepsilon}$ affects every channel. This is true
either if $\varepsilon$ acts randomly on a limited number of channels or when it affects all of them. Therefore,
the procedure improves the fidelity of transmission of quantum states only if the error in the QT channel is small,
that means $p_{\mu} < \frac{1}{2}$. Therefore, there should necessarily be some entanglement
in order to realize a useful protocol.


\section{Other examples}\label{sec6}

Among all the examples presented so far, the QT scheme plays a special role, as
the physical channel is not used at all. Moreover, QT aims at correcting any
noisy channel. In this section, we provide examples of both deterministic and probabilistic
protocols designed to reduce the complexity of a fixed noisy channel; these will also rely on the
physical channel for the transmission of the state.

We start by mentioning the simplest possible case: the error is given by a unitary shift of
the state,
\begin{equation}\label{errmod1}
    \rho^B = \varepsilon [\rho^A] = U_{\varepsilon} \rho^A U_{\varepsilon}^{\dagger},
\end{equation}
where $U_{\varepsilon}$ is a known, unitary operator. Since there is only one Kraus operator,
the complexity of the channel is $\chi (\varepsilon) = 1$, and the associated Choi-Jamiolkowski
state $R^{BA}$ is pure. The simplest strategy to deterministically correct this error does not
require ancillae or CC; one may just apply the local operations given by $U^A= U_{\varepsilon}^{\dagger}$ and
$U^B = I^B$, or $U^A = I^A$ and $U^B= U_{\varepsilon}^{\dagger}$. More generally, it is possible
to consider $U^A = U_1$ and $U^B = U_2$, with $U_1$, $U_2$ unitary operators such that $U_2 U_{\varepsilon}
U_1 = I^A$. It follows that $\lambda$ is a unitary transformation with a single Kraus operator
\begin{equation}\label{lambda0}
    \Lambda_0 = U_2 \otimes U_1^T.
\end{equation}
It is possible to check that $\lambda [R^{BA}] = \Psi_0^{BA}$, and then $\chi (\tilde{\varepsilon}) = 0$.

A more interesting example is the following. We consider the transmission of a qubit ($N = 2$)
subject to a bit-flip error,
\begin{equation}\label{errmod2}
    \rho^B = \varepsilon [\rho^A] = (1 - p) \rho^A + p \sigma_z \rho^A \sigma_z,
\end{equation}
and $p \in [0, 1]$. The Choi-Jamiolkowski state associated
to $\varepsilon$ is given by
\begin{equation}\label{chja2}
    R^{BA} = (1 - p) \Psi^{BA}_0 + p \Psi^{BA}_1,
\end{equation}
with non-vanishing eigenvalues $r_0 = 1-p$, and $r_1 = p$, and corresponding normalized eigenvectors
given by the Bell states
\begin{equation}\label{eigv2}
    \vert r_0 \rangle^{BA} = \vert \psi_0 \rangle^{BA}, \qquad
    \vert r_1 \rangle^{BA} = \vert \psi_1 \rangle^{BA},
\end{equation}
hence $\chi (\varepsilon) = 2$.

\begin{figure}[b]
\centering
\begin{picture}(140,130)(0,10)

\linethickness{0,35pt}
\put(-40,98){\line(1,0){24}}
\put(8,98){\line(1,0){42}}
\put(90,98){\line(1,0){14}}
\put(156,98){\line(1,0){24}}
\put(8,78){\line(1,0){4}}
\put(128,98){\line(1,0){4}}
\put(36,78){\line(1,0){68}}
\put(128,78){\line(1,0){4}}
\put(104,58){\line(1,0){28}}
\put(40,118){\vector(1,0){60}}
\put(50,90){\framebox(40,16)[]{$\varepsilon$}}
\put(-35,101){\makebox(0,0)[b]{$\rho^A$}}
\put(179,101){\makebox(0,0)[b]{$\tilde{\rho}^B$}}
\put(-20,22){\dashbox{1}(60,111)[]{}}
\put(100,22){\dashbox{1}(60,111)[]{}}
\put(70,65){\makebox(0,0)[b]{$\rho^{ab^{\prime}}$}}
\put(110,45){\makebox(0,0)[b]{$\rho^{b^{\prime \prime}}$}}
\put(69,122.5){\makebox(0,0)[b]{$CC$}}
\put(-35,92){\makebox(0,0)[b]{$\rightarrow$}}
\put(175,92){\makebox(0,0)[b]{$\rightarrow$}}

\put(-16,69){\dashbox{1}(24,39)[]{$U^{Aa}$}}
\put(12,69){\dashbox{1}(24,19)[]{$\Pi^a_{\eta}$}}
\put(104,69){\dashbox{1}(24,39)[]{$\bar{\Pi}_{\xi}^{Bb^{\prime}}$}}
\put(132,49){\dashbox{1}(24,59)[]{$\bar{U}_{\eta, \xi}^{Bb}$}}
\put(24,88){\vector(0,1){30}}
\put(116,108){\vector(0,1){10}}
\put(144,118){\vector(0,-1){10}}
\put(24,118){\line(1,0){16}}
\put(100,118){\line(1,0){44}}
\put(-8,27){\makebox(0,0)[b]{$L^{Aa}$}}
\put(112,27){\makebox(0,0)[b]{$L^{Bb}$}}

\end{picture}
\caption{Protocol used for the probabilistic correction of the bit-flip channel.}\label{fig4}
\end{figure}
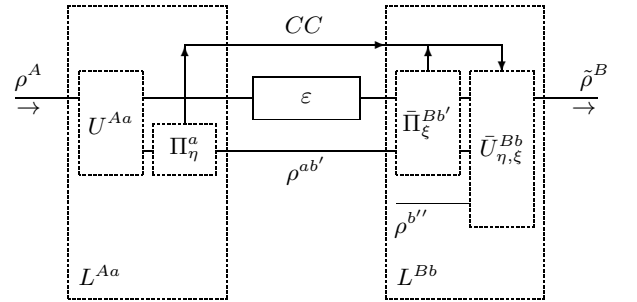

The protocol for reducing the complexity is based on the use of auxiliary systems, CC, and
local operations, following the scheme depicted in Fig. \ref{fig4}.
The ancilla is of the form $\rho^{ab} = \rho^{ab^{\prime}} \otimes \rho^{b^{\prime \prime}}$,
where ${\mathcal H}^b = {\mathcal H}^{b^{\prime}} \otimes {\mathcal H}^{b^{\prime \prime}}$. The systems
denoted by $a, b^{\prime}$ are qubits, with $\rho^{a b^{\prime}} = \vert \psi \rangle^{ab} \langle \psi \vert$
a non-maximally entangled pure state, where
\begin{equation}\label{shmiex2}
    \vert \psi \rangle^{ab} = \mu \vert 0 0 \rangle^{ab} + \sqrt{1 - \mu^2} \vert 1 1 \rangle^{ab},
\end{equation}
and Schmidt coefficients $\mu \leqslant \sqrt{1- \mu^2}$ without loss of generality. On the $A$ side, the local operations are a
unitary gate followed by a projective measurement of the ancilla, whose result is sent to $B$. This corresponds to
\begin{equation}\label{locA2}
    U^{Aa} = U_{cnot}^{Aa}, \quad \Pi_{\eta}^{Aa} = I^A \otimes \Pi_{\eta}^a = I^A \otimes \vert \eta \rangle^a \langle \eta \vert,
\end{equation}
with $\eta = 0, 1$, and $A, a$ playing the roles of control and target, respectively, for the CNOT gate. On the $B$
side, we have a projective measurement $\bar{\Pi}_{\xi}^{Bb}$, $\xi = 0, 1$, defined by
\begin{eqnarray}\label{locB2}
    \bar{\Pi}_0^{Bb} &=& (\vert 00 \rangle^{B b^{\prime}} \langle 00 \vert +
    \vert 11 \rangle^{B b^{\prime}} \langle 11 \vert) \otimes I^{b^{\prime \prime}}, \nonumber \\
    \bar{\Pi}_1^{Bb} &=& (\vert 01 \rangle^{B b^{\prime}} \langle 01 \vert +
    \vert 10 \rangle^{B b^{\prime}} \langle 10 \vert) \otimes I^{b^{\prime \prime}},
\end{eqnarray}
followed by a unitary operation of the form $\bar{U}_{\eta,\xi}^{Bb} = \bar{U}_{\eta,\xi}^{Bb^{\prime \prime}} \bar{U}_{cnot}^{B b^{\prime}}$,
with $B$ control, $b^{\prime}$ target, for the CNOT gate. The operator $\bar{U}_{\eta,\xi}^{Bb^{\prime \prime}}$ is used to realize two
POVMs defined by
\begin{eqnarray}\label{povm1}
    \varphi [\rho] &=& F_s \rho F_s + F_u \rho F_u,  \nonumber \\
    \varphi^{\prime} [\rho] &=& F^{\prime}_s \rho F^{\prime}_s + F^{\prime}_u \rho F^{\prime}_u,
\end{eqnarray}
with Kraus operators, associated to successful events, having the form
\begin{equation}\label{locfilB2}
    F_s = \left[
              \begin{array}{cc}
                1 & 0 \\
                0 & \gamma \\
              \end{array}
            \right], \,\,
    F_s^{\prime} = \left[
              \begin{array}{cc}
                \gamma & 0 \\
                0 & 1 \\
              \end{array}
            \right], \quad \gamma = \frac{\mu}{\sqrt{1 - \mu^2}}
\end{equation}
in the computational basis,
and inconclusive events described by $F_u, F_u^{\prime}$ such that $F_s^2 + F_u^2 = I^B$,
and analogously for primed quantities. These POVMs are applied following the prescriptions
of Table \ref{Tab1}. In other words, the receiver decides a probabilistic strategy of correction
based on both the measurement results $(\eta, \xi)$.
\begin{table}[t]
  \centering
  \begin{tabular}{|c|c|}
  \hline
  Values of $(\eta, \xi)$ & Strategy \\ \hline
  0, 0 & $\varphi$ \\
  0, 1 & $\varphi \circ \sigma_x$ \\
  1, 0 & $\varphi^{\prime} \circ \sigma_x$ \\
  1, 1 & $\varphi^{\prime}$ \\
  \hline
\end{tabular}
  \caption{Strategy to be followed by $B$ to probabilistically correct the error induced by the bit-flip
  channel, depending on the measurement result of both the projective measurements performed by $A$ and $B$}\label{Tab1}
\end{table}
The explicit form of $\bar{U}_{\eta,\xi}^{Bb^{\prime \prime}}$
and $\rho^{b^{\prime \prime}}$ is not important. An upper bar marks all the unitary and projection
operators acting on the $B$ side, to distinguish them from the operators appearing in (\ref{chan2}).
We find convenient to express $L^{Bb}$ in this form; therefore in all the formulas there
will be the upper bar and an additional index $\xi$. Accordingly and following (\ref{auxop}),
we can evaluate the operators $\bar{A}^{\eta}_{i, j}$ and $\bar{B}^{\eta, \xi}_{i, j}$. The only non-vanishing
terms are
\begin{eqnarray}\label{auxopex2}
    \bar{A}^0_{0,0} = \vert 0 \rangle^A \langle 0 \vert, &\quad& \bar{A}^0_{0,1} = \vert 1 \rangle^A \langle 1 \vert, \nonumber \\
    \bar{A}^1_{1,0} = \vert 1 \rangle^A \langle 1 \vert, &\quad& \bar{A}^1_{1,1} = \vert 0 \rangle^A \langle 0 \vert, \nonumber \\
    \bar{B}^{0,0}_{0,0} = \vert 0 \rangle^B \langle 0 \vert, &\quad& \bar{B}^{0,0}_{0,1} = \gamma \vert 1 \rangle^B \langle 1 \vert, \nonumber \\
    \bar{B}^{0,1}_{1,0} = \vert 0 \rangle^B \langle 1 \vert, &\quad& \bar{B}^{0,1}_{1,1} = \gamma \vert 1 \rangle^B \langle 0 \vert, \nonumber \\
    \bar{B}^{1,0}_{0,0} = \vert 1 \rangle^B \langle 0 \vert, &\quad& \bar{B}^{1,0}_{0,1} = \gamma \vert 0 \rangle^B \langle 1 \vert, \nonumber \\
    \bar{B}^{1,1}_{1,0} = \vert 1 \rangle^B \langle 1 \vert, &\quad& \bar{B}^{1,1}_{1,1} = \gamma \vert 0 \rangle^B \langle 0 \vert.
\end{eqnarray}
In agreement with (\ref{kraujan}), we derive the Kraus operators
\begin{eqnarray}
  \bar{\Lambda}^{0,0}_{0,0} &=& \mu (\vert 0 \rangle^B \langle 0 \vert \otimes \vert 0 \rangle^A \langle 0 \vert +
  \vert 1 \rangle^B \langle 1 \vert \otimes \vert 1 \rangle^A \langle 1 \vert), \nonumber \\
  \bar{\Lambda}^{0,1}_{1,0} &=& \mu (\vert 0 \rangle^B \langle 1 \vert \otimes \vert 0 \rangle^A \langle 0 \vert +
  \vert 1 \rangle^B \langle 0 \vert \otimes \vert 1 \rangle^A \langle 1 \vert), \nonumber \\
  \bar{\Lambda}^{1,1}_{1,1} &=& \bar{\Lambda}^{0,0}_{0,0}, \qquad
  \bar{\Lambda}^{1,0}_{0,1} = \bar{\Lambda}^{0,1}_{1,0}.
\end{eqnarray}
These operators satisfy
\begin{eqnarray}
  \bar{\Lambda}^{0,0}_{0,0} \vert r_0 \rangle^{BA} &=& \bar{\Lambda}^{1,1}_{1,1} \vert r_0 \rangle^{BA} = \mu \vert \psi_0 \rangle^{BA}, \nonumber \\
  \bar{\Lambda}^{0,0}_{0,0} \vert r_1 \rangle^{BA} &=& \bar{\Lambda}^{1,1}_{1,1} \vert r_1 \rangle^{BA} = 0, \nonumber \\
  \bar{\Lambda}^{0,1}_{1,0} \vert r_0 \rangle^{BA} &=& \bar{\Lambda}^{1,0}_{0,1} \vert r_0 \rangle^{BA} = 0, \nonumber \\
  \bar{\Lambda}^{0,1}_{1,0} \vert r_1 \rangle^{BA} &=& \bar{\Lambda}^{0,1}_{1,0} \vert r_1 \rangle^{BA} = \mu \vert \psi_0 \rangle^{BA},
\end{eqnarray}
therefore
\begin{equation}\label{correction2}
    \tilde{R}^{BA} = \sum_{\eta, \xi} \sum_{k, l} \bar{\Lambda}^{\eta, \xi}_{k, l} R^{BA}
    \bar{\Lambda}^{\eta, \xi \dagger}_{k, l} = 2\mu^2 \Psi_0^{BA},
\end{equation}
and the protocol reduces the complexity of the channel to $\chi (\tilde{\varepsilon}) = 0$ with
probability $2 \mu^2$. The protocol is probabilistic unless $\rho^{a b^{\prime}}$ is a maximally
entangled state. Notice that conclusive QT is a {\it different} approach to probabilistically correct this
channel {\it with the same success probability}. However, in this case, the physical channel
is not used at all, as usual in QT protocols.

\section{Discussion and conclusions}\label{sec8}

In this paper, we have described how to control a noisy channel from the point of
view of geometric control. The control actions are given by manipulations that can be locally performed by sender and
receiver, by the entanglement they share, and by classical communication they can use
to correlate their operations. It has been shown how each of these resources separately enter
the channel manipulation, by using the Choi-Jamiolkowski isomorphism between quantum
operations and non-normalized states. In particular, controlling a channel corresponds to
engineering the Kraus operators that characterize the map between Choi-Jamiolkowski states.
We have defined the notion of complexity of a noisy channel as a figure of merit for improving
the performance of the channel, and discussed its meaning and relevance. From this point of view,
we have described the roles of the standard schemes of QT and QECC.

In our analysis, the resources used for the channel manipulation play a prominent role. In fact,
for the feasible implementation of quantum technologies, extra resources represent a critical factor,
and protocols with reduced resources are, in general, preferred. Conversely, poor resources
(e.g., non-maximal shared entanglement) means a reduced ability in counteracting the noise.

The complexity represents a natural figure of merit when the geometric control approach is
used. From this point of view, the problem of simplifying the structure of the channel is equivalent to
the problem of properly driving the corresponding Choi-Jamiolkowski state. A reduction
of complexity leads to a simplified error processing of the channel, e.g., a QECC with reduced
resources for its implementation. Therefore, our work is of interest for models of adaptive QECC,
that is, the analysis of the minimal QECC to be used to perfectly correct a particular noisy channel.

We conclude by mentioning some possible generalizations of our work.
While the notion of complexity of the channel can be used as a guideline for the
application of our formalism, it is not intended to represent the only possible figure of merit.
In fact, the model developed in this paper constitutes the general connection between the investigation
of the properties of noisy channels and control theory. By choosing other figures of merit,
a different analysis can be worked out. For example, one could accept an imperfect
correction of the noisy channel and try to improve the fidelity of state transmission for a
given channel and a fixed noise model. In this case, the fidelity is the relevant quantity, and methods
and ideas coming from the theory of optimal control could be directly applied.

Another generalization of our work has already been mentioned in the Introduction.
The formalism presented in this paper can be directly applied to the
general case of quantum processing, and it is not limited to quantum communication.
In fact, the ideal operation $\varepsilon_0$ can be an arbitrary unitary operation, rather
than the identity ${\mathcal I}$. The corresponding Choi-Jamiolkowski state is an arbitrary
pure state (rather than the Bell state $\Psi_0$), and the complete formalism perfectly
adapts to this case.


In our approach, the control of the noisy channel $\varepsilon$ is expressed in terms of
its Choi-Jamiolkowski state $R^{BA}$, by means of constant control parameters $u$. This is due to the
implicit assumption that time does not directly enter our analysis: the noisy channel is
fixed, and the local operations $L^{Aa}$ and $L^{Bb}$ can be performed almost instantaneously.
These actions precede or follow the transmission of the quantum state through the physical channel.

However, under some circumstances, these assumptions could be inappropriate; for instance,
when the transmission time is not fixed, or when the local operations are not instantaneous
and well separated in time. Operations distributed over time (e.g., a continuous measurement)
could overlap in a non-trivial way, and would not fit into the previous description.
In control theory, it is common to express the evolution of a system (represented by a
generic variable $x$) in differential form, and to admit controls that are functions of time, $u = u(t)$,
\begin{equation}\label{gencon}
    \dot{x}(t) = f(x(t), u(t), t),
\end{equation}
where $f$ is a suitable vector field. This scenario can be used also in the
control analysis of noisy channels, to deal with the aforementioned situations. The basic
requirements are $x (0) = \Psi_0 \simeq {\mathcal I}$ and $x (\tau) = R^{BA} \simeq \varepsilon$ in the absence
of control, with $\tau$ the standard transmission time. In this case, (\ref{gencon}) is a Master Equation
accounting for the decohering action of the environment as well as the control actions which, in general,
can be modulated in time. While the derivation of this equation is beyond the scope of this paper,
we mention that the case of coherent control of the time propagator of a closed system
has already been considered in the context of geometric control theory.

Finally, we want to mention that, in this paper, we have considered the transmission of an $N$-level quantum system through a noisy
channel, with $N$ arbitrary but finite. Our formalism may also be used in the case of
continuous variable systems, corresponding to $N \rightarrow \infty$. In fact, the Choi-Jamiolkowski
isomorphism is still valid in this regime, even if some care has to be taken in order to properly
define physically implementable, maximally entangled states~\cite{fiur,gied}.
In particular, Gaussian states and operations fit well into our framework. The notion of complexity of the noisy
channel is still meaningful, even though, in this case, this quantity can be arbitrarily large. In fact, in the
analysis of continuous variable systems, the Kraus operators can depend on a continuous variable
rather than a discrete one.


\appendix
\section{Upper bound on the number of Kraus operators}\label{app1}

An arbitrary completely positive map $\lambda$ can be expressed in the operator sum representation
by using several sets of Kraus operators, according to the transformation $\Lambda_i \rightarrow
\sum_i u_{ij} \Lambda_j$. The cardinalities of these sets are in general different. The aim of
this appendix is to derive the minimal number of Kraus operators that is needed for a specific map.
By means of it, we can derive an upper bound on the irreducible number of Kraus operators for a
dynamical evolution that satisfies a given set of constraints (the imposed transitions).

We refer to the scenario previously discussed, where $\lambda$
acts on states in ${\mathcal H}_B \otimes {\mathcal H}_A$ and, in principle, it may
contain at most $N^4$ Kraus operators, since the space has dimension $N^2$.
For convenience of notation, we denote by $R$ the rank of $R^{BA}$;
this is the number of constraints the dynamics must satisfy. We have shown previously that at least $R$
operators should appear in the operator sum representation of $\lambda$.
In the Kraus operators $\Lambda_j$, the index $j$ may denote a triple $(\eta, k, l)$
or a pair $(k, l)$, depending on the case at hand (with or without CC, respectively).

We introduce two orthonormal bases for ${\mathcal H}_B \otimes {\mathcal H}_A$, ${\mathcal B}
= \{ \vert b_i \rangle; i\}$ and ${\mathcal B}^{\prime} = \{ \vert b_i^{\prime} \rangle; i\}$,
with $i = 0, \ldots, N^2 - 1 $, such that $\vert b_i \rangle = \vert r_i \rangle^{BA}$ for $i = 0, \ldots R - 1$, and
$\vert b^{\prime}_0 \rangle = \vert \psi_0 \rangle^{BA}$. With this choice of the bases, the needed
Kraus operators are represented by the $N^2 \times N^2$ matrices
\begin{equation}\label{lambda1}
    [\Lambda_j]_{{\mathcal B} {\mathcal B}^{\prime}} = \left[
                    \begin{array}{cc}
                      \vec{\delta}_j & \vec{0} \\
                      \mathcal{O} & M_j \\
                    \end{array}
                  \right], \quad j = 1, \ldots, R,
\end{equation}
where $\vec{\delta}_j$ is an $R$-dimensional row vector satisfying $(\vec{\delta}_j)_i = \delta_{ij}$,
$\mathcal{O}$ is the $(N^2 - 1) \times R$ null matrix, $\vec{0}$ is the $(N^2 - R)$-dimensional null
row vector, and $M_j$ a $(N^2 - 1) \times (N^2 - R)$ matrix. Additional $r$ Kraus operators have the form
\begin{equation}\label{lambda2}
    [\Lambda_j]_{{\mathcal B} {\mathcal B}^{\prime}} = \left[
                    \begin{array}{cc}
                      \mathcal{O} & M_j \\
                    \end{array}
                  \right], \quad j = R + 1, \ldots, R + r,
\end{equation}
where $\mathcal{O}$ is the $N^2 \times R$ null matrix,
and $M_j$ a $N^2 \times (N^2 - R)$ matrix. Condition (\ref{trajan2}) reads
\begin{equation}\label{traprem}
    \sum_{j = 1}^{R + r} M^{\dagger}_j M_j \leqslant ({\rm dim} {\mathcal H}_a) I^{BA}.
\end{equation}
The freedom in engineering the map $\lambda$, that realize the desired task, amounts to
different choices of $r$, and of the corresponding Kraus operators. We want to find the largest
$r$ such that there is not an equivalent representation of $\lambda$ in terms of a
reduced number of Kraus operators. This $r$ will be a measure of the aforementioned freedom.

Using the Gauss elimination procedure, it is possible to prove that if $r > N^2 (N^2 - R)$ there
always exist $R + r$ complex coefficients $\gamma_i$ such that
\begin{equation}\label{vinc}
    \sum_{j = 1}^{R + r} \gamma_j \Lambda_j = 0, \qquad \sum_{j = 1}^{R + r} \vert \gamma_j \vert^2 =1.
\end{equation}
But these coefficients can be used as a row of the unitary matrix that defines an equivalent
set of Kraus operators, $u_{ij} = \gamma_j$ for an arbitrary $i$. Because of (\ref{vinc}), the
corresponding map $\lambda$ can be represented by using $R + r - 1$ Kraus operators. Therefore,
the largest $r$ we are looking for is given by $r = N^2 (N^2 - R)$. We conclude that the upper bound
for the number of Kraus operators is given by $N^4 - R (N^2 - 1)$.


\end{document}